\documentclass[%
 reprint,
superscriptaddress,
 amsmath,amssymb,
 aps,
]{revtex4-2}

\usepackage{xcolor}
\usepackage{amsmath,amssymb,amsfonts}
\usepackage{graphicx}
\usepackage{textcomp,nicefrac}
\usepackage{tabularx}
\usepackage[normalem]{ulem}
\usepackage{enumitem}

\begin{document}

\title{A Bayesian Optimization Approach for Attenuation Correction \\ in SPECT Brain Imaging}

\author{Loizos Koutsantonis}%
\email{loizos.koutsantonis@uni.lu}
\author{Ayman Makki}
\email{ayman.makki@uni.lu}
\author{Tiago Carneiro}
\email{tiago.carneiropessoa@uni.lu}
\author{Emmanuel Kieffer}
\email{emmanuel.kieffer@uni.lu}
\author{Pascal Bouvry}
\email{pascal.bouvry@uni.lu}

\affiliation{Department of Computer Science, Faculty of Science, Technology and Medicine, University of Luxembourg, Luxembourg}


\date{\today}
\begin{abstract}
 Photon attenuation and scatter are the two  main physical factors affecting the diagnostic quality of SPECT in its applications in brain imaging. 
In this work, we present a novel Bayesian Optimization approach for Attenuation Correction (BOAC) in SPECT brain imaging. BOAC utilizes a prior model parametrizing the head geometry and exploits High Performance Computing (HPC) to reconstruct attenuation corrected images without requiring prior anatomical information from complementary CT scans. BOAC is demonstrated in SPECT brain imaging using noisy and attenuated sinograms, simulated from numerical phantoms. The quality of the tomographic images obtained with the proposed method are compared to those obtained without attenuation correction by employing the appropriate  image quality metrics.  The quantitative results show the capacity of BOAC  to provide  images exhibiting  higher contrast and less background artifacts as compared to the non-attenuation corrected MLEM images.  
 
\end{abstract}

\maketitle

\section{Introduction}

Single Photon Emission Computed Tomography (SPECT) allowing the visualization of physiological processes within the human body is being used in brain imaging studies for the diagnosis and monitoring of various brain disorders. Clinical applications of brain SPECT imaging include the localization of brain tumors with $^{201}$Tl and the dopamine transporter imaging with $^{123}$I for the diagnosis of Parkinson's disease.

The diagnostic capacity of SPECT is greatly influenced by the quality of the obtained tomographic images. In particular, for the case of brain imaging, the varying photon attenuation and scatter in the inhomogeneous tissue structure of the head (grey/white matter, cranium) affects the accuracy and specificity of  SPECT imaging. So far, various techniques have been developed to compensate for the effects of attenuation and scatter in the reconstructed images. These techniques use an attenuation map to construct the system matrix of the tomographic problem and provide a forward model that accounts for the attenuation of photons in the imaged region \cite{Cam, Zei, tavakoli}.  With the advent of hybrid SPECT/CT devices,  patient-specific attenuation maps could be generated using the available anatomical information from the complementary CT scan and used for the reconstruction of SPECT images \cite{Shim, Lapa, Bie, Sien}. While this approach can provide spatially varying attenuation maps, it exposes the patient to increased radiation doses \cite{Sharma}.

In this work, we present a novel approach for reconstructing attenuation corrected images in brain imaging studies without using anatomical information from complementary CT scans. The technique uses a prior model parametrizing the distribution of attenuation coefficients within the head and employs the Maximum Likelihood Expectation Maximization Method (MLEM) in a Bayesian optimization framework to define the optimum attenuation map from the obtained data (sinogram). This map is used in MLEM to reconstruct attenuation corrected images.

\section{Attenuation Model}

The attenuated sinogram data obtained in emission tomography (SPECT, PET) can be described using  the attenuated Radon transform \cite{Natterer}:
\begin{equation}
     S(\phi,s) = \int _{t(s,\phi)} f(x,y) e^{-\int_{D(s,\phi)} \mu(x,y) dD} dt
     \label{eq:Radon}
\end{equation}
 where, $S(\phi, s)$ is the bin measurement obtained at the offset $s$  and the $\gamma$-camera's rotation angle $ \phi $, $f(x,y)$ is the 2D activity distribution  within the tomographic plane  and $\mu(x,y)$ is the 2D distribution of attenuation coefficients of the imaged structure. In the case where prior anatomical information is available from a CT scan, the distribution $\mu(x,y)$ can be estimated and used to compose the system matrix $A^{\mu}$ linking the sinogram data $S$ to the vectorized representation $F$ of the 2D activity distribution $f(x,y)$ . 
 \begin{equation}
     S = A^{\mu} F
     \label{Eq:Cost}
\end{equation}

\begin{figure}[t!]
\centering
\includegraphics[width = 1.0\columnwidth]{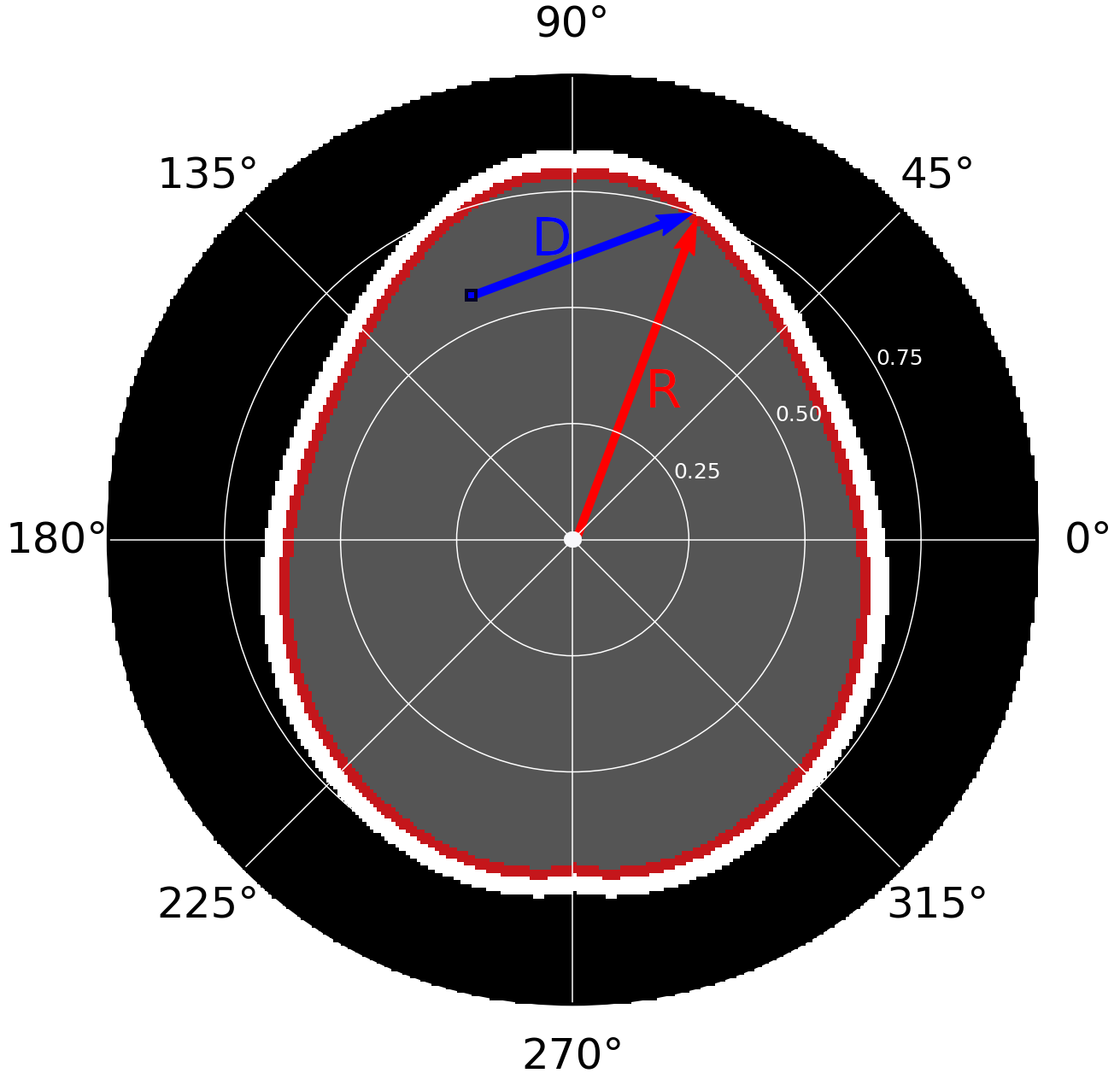}
\caption{Polar plot depicting the model representation of the attenuation map in the proposed BOAC method. The head geometry is described by a truncated series of Legendre polynomials and their corresponding coefficients. The optimum coefficients' values for a given set of measurements (sinogram)  are computed in a Bayesian Optimization scheme to derive the system matrix of the SPECT tomographic problem.}
\label{fig:model}
\end{figure}

 In the approach presented in this work, the geometry of the imaged structure (head) is modeled using a truncated series of Legendre polynomials:
 \begin{equation}
    R(\theta) = \sum_{k=0)}^m c_k P_k(\theta) 
    \label{eq:Geometry}.
 \end{equation}
 where,  $P_k$ is the set of first six Legendre polynomials and $c_k$ their corresponding weighting  coefficients. This parametrization of the head geometry, also depicted in Fig. \ref{fig:model}, is used to construct a model-based attenuation map $\mu(r, \theta)$ :
\begin{equation}
\mu(r, \theta) =
\begin{cases} 
\mu _b, & r<R(\theta)\\ 
\mu _c, &  R(\theta)<r<R(\theta)+\delta R\\
0, & R(\theta)+\delta R<r
\end{cases} 
\label{eq:attenuation}
\end{equation}
where, $\mu _b$ and $\mu _c$ are the attenuation coefficients of the brain matter and cranium respectively.

Given the above parametrization of the attenuation map, the image reconstruction problem becomes the problem of defining  the optimum coefficients' values $c^{opt}_k$ directly from  the emission data $S$. The tomographic image of the activity distribution $F$ is reconstructed using the system matrix $A^{\mu}$ corresponding to the optimum coefficients' values $c^{opt}_k$.

\section{Bayesian Optimization}

In SPECT, given the attenuation map $A_\mu$ and the activity distribution  $F$, the likelihood of measuring a sinogram $S$  is given by the Poisson distribution $P(S|F, A_{\mu})$. The log-likelihood function for the Poisson distribution is given by:
\begin{equation}
     \ln  P(S|F, A_{\mu})  = \sum_i \bigg( S_i\ln\sum_j A^{\mu}_{ij}F_j - \sum_j A^{\mu}_{ij} F_j - \ln S_j! \bigg)
     \label{eq:log}
\end{equation}
The above formula provides a measure of linkage between the sinogram data $S$ and the unknown parameters  $c_k$ and $F$.  Assuming that no prior information is available, and thus $P(S|F, A_{\mu}) \sim P(F, A_{\mu}|S)$,  the optimum coefficients' values $c^{opt}_k$  and tomographic image $\hat F$ best describing the sinogram can be computed by minimizing the negative log-likelihood function:
\begin{equation}
     \hat F, c^{opt}_k = \arg\min _{F, c_k} \bigg(-\ln  P(S|F, A_{\mu}(c_k)) \bigg)
\end{equation}

The solution proposed in this work implements the Maximum Likelihood Expectation Maximization (MLEM) \cite{Shepp} method using Bayesian Optimization (BO) \cite{Pelikan} scheme to determine the optimum coefficients' values  $c^{opt}_k$ and reconstruct the tomographic image $\hat F$. BO is a model-based approach with the sole purpose of tackling time-consuming objective functions. BO has become an essential and top-notch approach to solve many real-world applications in fields such as combinatorial optimisation \cite{bo_combina}, robotics \cite{bo_robotic}, sensor networks \cite{bo_sensor}, environmental monitoring \cite{bo_env} and automated machine learning \cite{bo_auto}. This work is another application case in which BO demonstrates its capacity to tackle complex and time-consuming objective functions.    

To reduce the inherent computation cost, BO relies on a surrogate model of the true log-likelihood function which is sequentially refined by sampling new promising solutions. Sampling is achieved with a dedicated ``acquisition function'' aiming at guiding how the parameters' space of the problem should be explored. BO can be therefore characterized by two main features:
\begin{itemize}
\item  The surrogate model of the log-likelihood function which, in this study, is built on Gaussian process (GP) regression:
\begin{equation}
    -\ln  P(A_{\mu}(c_k)) \sim \mathcal{GP}(\mu(c_k),k(c_k,c_k'))
\end{equation}

\item The class of acquisition function defining a strategy for the efficient sampling of new candidate points in the parameters' space.
\end{itemize}

In this work, the classical but proven Radial Basis Function (RBF)   with length-scale parameter  $l$ and  variance $\sigma_{0}^{2}$ has been chosen for kernel in the Gaussian Process Regression:
\begin{equation}
  k(c_k,c_k') = l \cdot \exp{ \left(-\frac{\|c_k-c_k'\|^2}{2\sigma_0^2}\right)}
\end{equation}

The Expected Improvement (EI) has been selected as the acquisition function to sample the new candidate points:
\begin{equation}
    EI(c_k)=\sigma(c_k)(\gamma(c_k)\,\Phi(\gamma(c_k)) + \phi(\gamma(c_k)))
\end{equation}
 where $\Phi$ is the standard cumulative distribution function and $\phi$ the standard normal probability density function. The function $\gamma(c_k)$ is given by:
 \begin{equation}
    \gamma(c_k) = \frac{\mu(c_k) + \ln P(A_{\mu}(c^{best}_k)) - \xi}{\sigma(c_k)}
 \end{equation}
  where $\xi$ is a parameter allowing to balance  exploration-exploitation.  
  The landscape of the acquisition function can be highly non-linear. The traditional local search algorithms implemented historically in BO have been replaced in this work by the Chaotic Optimization algorithm described in \cite{talbi} to cope with this issue.

\begin{figure}[t!]
\centering
\includegraphics[width = 1.0\columnwidth]{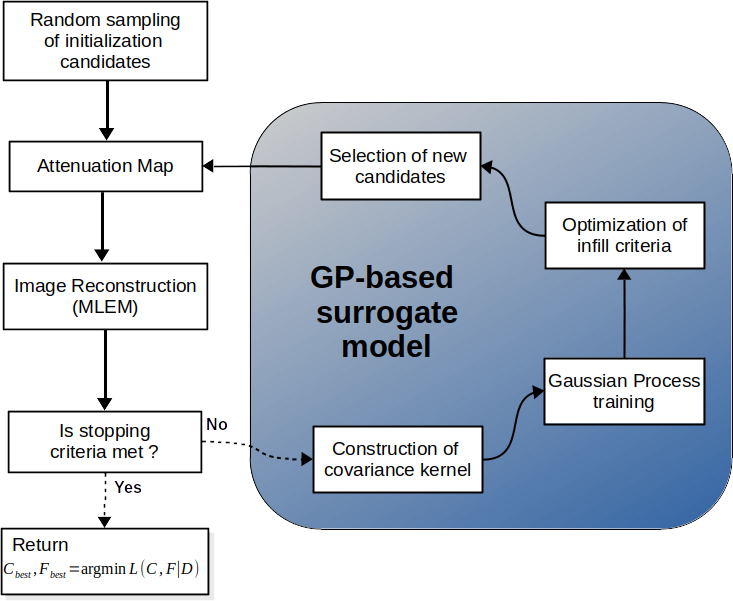}
\caption{Flowchart  of the  proposed Bayesian Optimization framework   for attenuation correction in SPECT brain imaging. }
\label{fig:flowchart}
\end{figure}

The BOAC framework is presented in the flowchart of Fig. \ref{fig:flowchart};   the implementation of the method is described by the following steps:
\begin{enumerate}
    \item An initial set of candidates $c_k$ is randomly chosen in a defined space. \item The geometry $R(\theta)$ of the imaged structure is calculated for each candidate set $c_k$ from Eq. \ref{eq:Geometry} and used in Eq. \ref{eq:attenuation} to compose the corresponding attenuation map $\mu(r, \theta)$ in polar coordinates.
    \item The system matrix of the tomographic problem is calculated for each candidate set of coefficients $c_k$ through Eq. \ref{eq:Radon}. A fixed number of MLEM iterations is used to reconstruct an image for each candidate set $c_k$ using the corresponding system matrix. A "goodness of fit" is assigned to each reconstructed image through the log-likelihood function (Eq. \ref{eq:log}).
    \item The available samples and their corresponding log-likelihood values are used in a Gaussian process regression to construct the surrogate model of the log-likelihood landscape.
    \item New candidate samples are obtained by optimizing an acquisition function over the surrogate model. For the optimization of the acquisition function, we used the chaotic optimization algorithm described in \cite{talbi}. The updated set of samples is used to initiate a new cycle of iterations. 
    \item The EI score of the "best" point is recorded at each cycle of iterations. The algorithm terminates when the EI score is below a specified threshold. 
\end{enumerate}

The calculation of the system matrix for a set of coefficients $c_k$ at each cycle of  BOAC requires a large amount of computation.
For the practical application of the method, the computation of the system matrix was parallelized on a NVIDIA Graphics Processing Unit (GPU) using the Compute Unified Device Architecture (CUDA). The forward-projection and back-projection operators of the MLEM iterative algorithm were also parallelized with CUDA to accelerate the image reconstruction process.

\begin{figure*}[ht!]
\centering
\includegraphics[width = \textwidth]{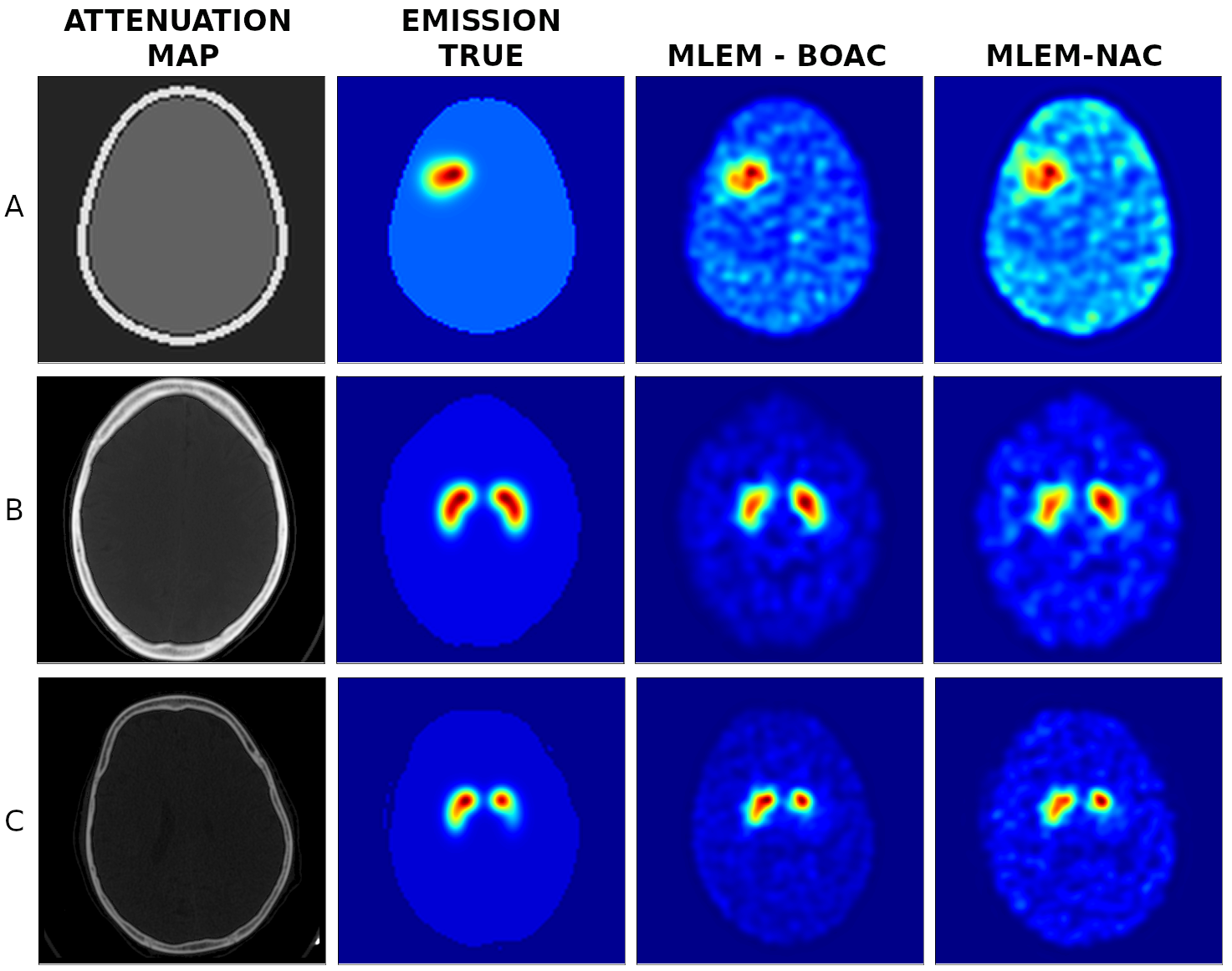}
\caption{ Reconstructed images obtained with the proposed method (MLEM-BOAC) and without attenuation correction (MLEM-NAC). The attenuation maps and the true emission distributions used for the simulation of the projection data are shown in first and second column respectively. }
\label{fig:recs}
\end{figure*}

\section{Simulation Studies}

The proposed method is demonstrated in SPECT brain imaging using data from numerical phantom simulations.   In three case studies, attenuated sinograms containing 72 projections uniformly distributed in the full angular range were simulated using the emission and attenuation images presented in Fig. \ref{fig:recs}. Emission and attenuation maps were defined on $128 \times 128$ grids. The simulated sinograms were further randomized with Poisson noise. The images reconstructed with BOAC are shown in Fig. \ref{fig:recs}. Reconstructions produced without attenuation correction (NAC) are shown in the same figure for comparison. Visually, BOAC images present less background noise and higher contrast as compared to the non-attenuation corrected images.

Quantitative evaluations of the reconstructed images obtained with BOAC  were performed against the reference NAC images.
The  Structural  Similarity  Index  (SSIM) was used to quantify the similarity between the reconstructed and ground truth emission images \cite{Wang}. The Contrast-to-Noise Ratio (CNR) was used to measure the contrast between the region of interest and the background. The Peak Signal-to-Noise Ratio (PSNR) was employed to provide a measure of the ratio between the maximum activity value and the amount of noise affecting the quality of the reconstructed images. Table \ref{tab:metrics} summarizes the results of quantitative metrics. BOAC images exhibit higher similarity to the ground truth images as indicated by the SSIM scores. The CNR and PSNR scores validated the visual observations of higher contrast and less noise presented in the BOAC images as compared to the NAC images.

\begin{table*}[t!]
\centering
\renewcommand{\arraystretch}{2.0}
\caption{SSIM, CNR and PSNR scores quantifying the quality of the reconstructed images presented in Fig. \ref{fig:recs}. }
\label{table}
\begin{tabularx}{0.9\textwidth}{>{\raggedright\arraybackslash}p{20pt} >{\centering\arraybackslash}p{80pt} >{\raggedleft\arraybackslash}X >{\raggedleft\arraybackslash}X >{\raggedleft\arraybackslash}X }
\hline\hline
Case & Method & SSIM & CNR & PSNR \\
\hline
A & MLEM-BOAC & 0.87  & 8.64 & 26.25 \\
  & MLEM-NAC & 0.72  & 6.33 & 3.63 \\
  \hline
B & MLEM-BOAC & 0.90 & 11.85 & 28.16\\
  & MLEM-NAC & 0.73 & 8.49 & 3.65\\
  \hline
C & MLEM-BOAC & 0.94 & 13.31 & 31.05\\
  & MLEM-NAC & 0.73 & 9.15 & 4.61\\

\hline\hline
\end{tabularx}
\label{tab:metrics}
\end{table*}

\clearpage

\section{Conclusions}

In this work, the BOAC framework is presented for attenuation correction in SPECT brain imaging. The proposed method uses a prior geometrical model to parametrize the attenuation map of the imaged structure and employs the MLEM to provide reconstruction results. In the proposed framework, the parameters defining the attenuation map are estimated in a Bayesian Optimization scheme directly from the emission data (sinogram) and used to calculate the system matrix of the tomographic problem.  The method allows the incorporation of attenuation correction within the SPECT image reconstruction without the need for any anatomical image from a structural modality (e.g. CT).

The method is demonstrated with sinogram data from numerical phantom simulations. In three distinct simulation cases, BOAC led to high contrast images characterized by less noisy artifacts compared to the non-attenuation corrected reconstructions.
Further experimentation with real data is required for the evaluation of the method in clinical studies.

\begin{acknowledgments}
The authors would like to thank the NVIDIA AI Technology Center Luxembourg for their positive comments and careful review of the current study.
\end{acknowledgments}

\end{document}